\documentclass[prb,superscriptaddress,twocolumn,notitlepage,showpacs,longbibliography]{revtex4-2}
\usepackage{float,subfig,xcolor}
\usepackage{amsmath}
\usepackage{amssymb}
\usepackage{graphicx}
\usepackage{caption}
\usepackage{txfonts}
\usepackage{mathtools}

\newcommand{\rrangle}{\rangle\rangle}

\begin{document}

\title{Quantum Fluctuation-Dissipation Theorem Far From Equilibrium}

\author{Zhedong Zhang}
\email{zzhan26@cityu.edu.hk}
\affiliation{Department of Physics, City University of Hong Kong, Kowloon, Hong Kong SAR}

\author{Xuanhua Wang}
\affiliation{Department of Physics and Astronomy, State University of New York at Stony Brook, Stony Brook, New York 11794, USA}

\author{Jin Wang}
\email{jin.wang.1@stonybrook.edu}
\affiliation{Department of Physics and Astronomy, State University of New York at Stony Brook, Stony Brook, New York 11794, USA}
\affiliation{Department of Chemistry, State University of New York at Stony Brook, Stony Brook, New York 11794, USA}

\thanks{ZDZ and XW contribute equally to this work}

\date{\today}

\begin{abstract}
 Fluctuations associated with relaxations in far-from-equilibrium regime is of fundamental interest for a large variety of systems within broad scales. Recent advances in techniques such as spectroscopy have generated the possibility for measuring the fluctuations of the mesoscopic systems in connection to the relaxation processes when driving the underlying quantum systems far from equilibrium. We present a general nonequilibrium Fluctuation-Dissipation Theorem (FDT) for quantum Markovian processes where the detailed-balance condition is violated. Apart from the fluctuations, the relaxation involves extra correlation that is governed by the quantum curl flux emerged in the far-from-equilibrium regime. Such a contribution vanishes for the thermal equilibrium, so that the conventional FDT is recovered. We finally apply the nonequilibrium FDT to the molecular junctions, elaborating the detailed-balance-breaking effects on the optical transmission spectrum. Our results have the advantage of and exceed the scope of the fluctuation-dissipation relation in the perturbative and near equilibrium regimes, and is of broad interest for the study of quantum thermodynamics.
\end{abstract}

\flushbottom
\maketitle
%
%
\thispagestyle{empty}


\section*{Introduction}
Quantum thermodynamics is an active subject of statistical mechanics emergent from quantum mechanics. Recent exciting advance raised subtle questions about the fluctuations related to heat and work in the nano-scaled systems driven far from equilibrium \cite{rossnagel2016single,brandao2015second,krishnamurthy2016micrometre,ghosh2018two}. The heat dissipation is linked to the fluctuations, intrinsically through the fluctuation-dissipation theorem. However, the traditional fluctuation-dissipation relation only holds for the quantum systems either at thermal equilibrium where the detailed balance is strictly obeyed or being weakly deviated from equilibrium \cite{callen1951irreversibility,kubo1957statistical,zwanzig2001nonequilibrium}. The extension of a similar relation to the far-from-equilibrium regime is urgent now, as desired by the progress of quantum thermodynamics \cite{prigogine1971biological,niedenzu2018quantum,zhang2019quantum,zhang2016fluctuation}. This is however an open issue such that an underlying microscopic theory is still lacking.

The fluctuation is a fundamental subject of the study in the fields broadly ranging from statistical mechanics and spectroscopy to economy and social sciences \cite{zhang2019quantum,korepin1997quantum,mukamel1999principles,hu2019machine}. The development of conventional statistical mechanics has enriched the study of fluctuations for the systems at finite temperature, which provided deeper insights for understanding the critical phenomena and phase transitions \cite{binney1992theory,wilson1975renormalization}. For those systems slightly deviated from thermal equilibrium, the fluctuations have been exploited using the perturbative method that leads to the thermal (mass) transport near the equilibrium as a result of thermal and electric conductivity \cite{kubo1957statistical,callen1948application,benzi1992lattice}. Despite these progresses, the study of fluctuations of the systems driven far from equilibrium is still far from complete. Recent technical advance made it accessible to measure the forward and backward transitions of the particles for Markovian quantum dynamics that leads to the Fluctuation Theorem (FT) at the microscopic level \cite{esposito2009nonequilibrium,esposito2007semiconductors,aaberg2018fully,andrieux2009fluctuation}. There are various degree of contexts to express the FT, e.g., Crook relation emphasizing irreversible work fluctuation \cite{crooks1999entropy,collin2005verification}, entropy fluctuation in both close and open systems \cite{seifert2005entropy,evans1993probability,crooks2000path,crooks1998nonequilibrium,seifert_RPP2012}. As an important result, the Jarzynski equality naturally follows the FTs for both classical and quantum systems \cite{jarzynski1997nonequilibrium,mukamel2003quantum,bartolotta2018jarzynski}, and has been validated in recent experiments \cite{liphardt2002equilibrium,hoang2018experimental}. Nevertheless, the FTs can reduce to the fluctuation-dissipation relation close to equilibrium, retrieving the Green-Kubo formula for transport coefficient \cite{bochkov1979fluctuation}. This yields the FDT in near-equilibrium regime. However, the connection of the fluctuation to dissipation in far-from-equilibrium regime is still elusive, which needs to be found out explicitly. This desires the approaches going beyond the perturbative treatment in stochastic dynamics, e.g., Boltzmann equation. For the classical systems far from equilibrium, the FDT have been developed using the Fokker-Planck equation for the stochastic dynamics \cite{feng2011potential,fang2019nonequilibrium,harada2005equality,Yllanes_PNAS2017}. For the nonequilibrium quantum systems as a counterpart of classical systems, the fluctuation-dissipation relation is an open issue.

The nonequilibrium density matrix plays the key role for the purpose, as the microscopic dynamics breaking the detailed balance is involved. The propagation of the density matrix bridges the gap between the microscopic dynamics and statistics, responsible for the relaxation and the spectroscopic line shape \cite{tanimura2006stochastic,zhang2019polariton,breuer2002theory}. 
The corresponding equations of motion giving the underlying laws which the ensemble obeys lay the foundation of the nonequilibrium statistical mechanics. This is on equal footing with the ensemble theory under equilibrium assumption which leads to the grand canonical ensemble. In this work, we develop a universal quantum mechanical fluctuation-dissipation relation for the steady states far from equilibrium, via the curl flux theory quantifying the detailed-balance-violation \cite{fang2019nonequilibrium,zhang2015landscape,zhang2017nonequilibrium,wang2008potential}.
The quantum curl flux can enable us to calculate the far-from-equilibrium distributions of the population and the coherence, beyond the specific models. These may provide the fundamental building blocks for nonequilibruim quantum statistical mechanics. Generically differing from the conventional FDT, our results elaborate an extra term dictated by the curl flux in the response function in terms of the fluctuations \cite{callen1948application,bochkov1979fluctuation,agarwal1972fluctuation}. This prominently characterizes the nonequilibrium nature due to the violation of detailed balance. We further calculate the linear transmission spectrum of a molecular junction model as a typical nonequilibrium quantum system that can be measured in the spectroscopic experiments nowadays.

\begin{figure*}[t]
 \captionsetup{justification=raggedright,singlelinecheck=false}
 \centering
   \includegraphics[scale=0.4]{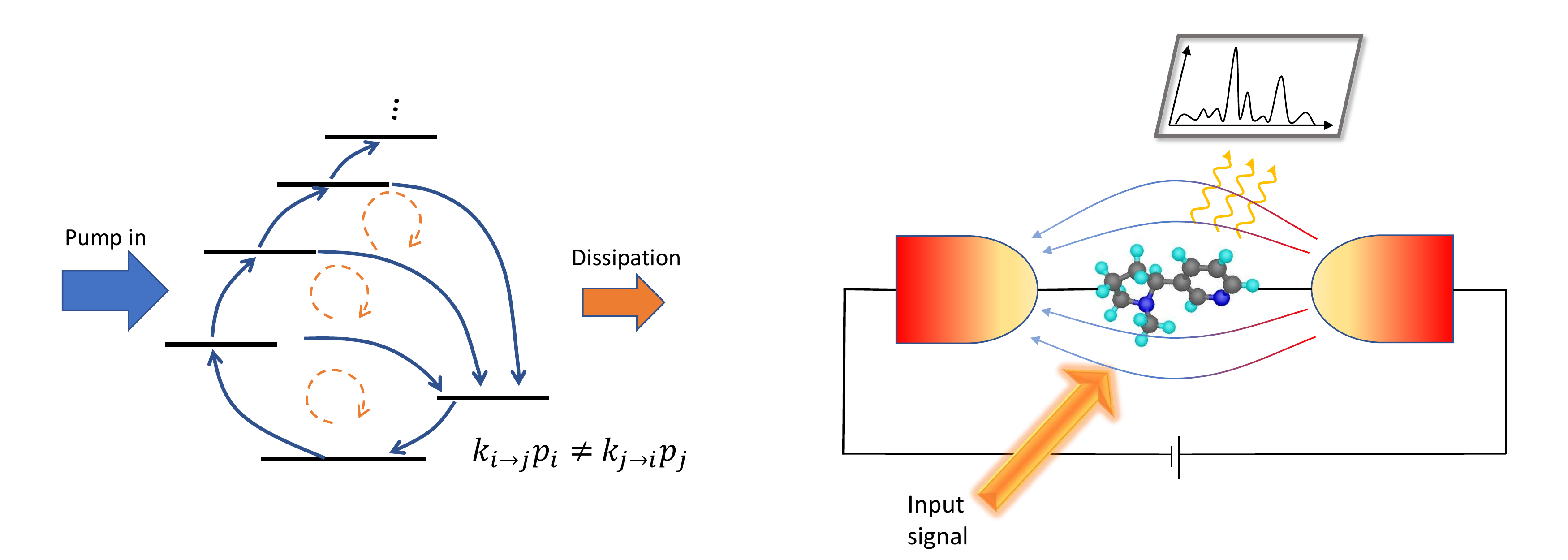}
\caption{(Left) Multilevel quantum system subject to energy pumping and dissipation. $k_{i\rightarrow j}$ is the transition rate from the $i^{th}$ level to the $j^{th}$ level and $p_i$ is the probability of the system at $i^{th}$ level. A group of loop fluxes exist as a result of detailed-balance-breaking. The magnitude and statistics of these loops quantify at microscopic level the detailed-balance-breaking. (Right) Optical transmission of a molecular junction driven far from equilibrium in the presence of large voltage bias. The role of electrodes is to exchange electrons with the molecules in between, so that the electric current is a result of the loop fluxes.} 
\label{Schm}
\end{figure*}

\section*{Curl Quantum Flux Decomposition}

The quantum dynamics of systems such as molecules can be described by quantum dissipative equation of motion, in the presence of energy dissipation and dephasing from the surrounding environments, such as phonons, low-frequency vibrations and solvent. Recalling the Born-Markov approximation applicable for smooth spectral density of the environments, the equation of motion (EOM) is of the most general form
\begin{equation}
    \begin{split}
     \frac{\partial}{\partial t}
        \begin{pmatrix}
           \rho_{\text{p}}\\[0.15cm]
           \rho_{\text{c}}
        \end{pmatrix}
      = \begin{pmatrix}
           M_{\text{p}} &  M_{\text{pc}}\\[0.15cm]
           M_{\text{cp}} & M_{\text{c}}
        \end{pmatrix}
        \begin{pmatrix}
           \rho_{\text{p}}\\[0.15cm]
           \rho_{\text{c}}
        \end{pmatrix}
    \end{split}
\label{qme1}
\end{equation}
where the density matrix is partitioned into population and coherence components in Liouville space, i.e.,  $|\rho\rangle\rangle=|\rho_{\text{p}}\rangle\rangle\oplus |\rho_{\text{c}}\rangle\rangle$ and the inner product is defined as $\langle\langle A|B\rangle\rangle=\text{Tr}(A^{\dagger}B)$. Notice that Eq.(\ref{qme1}) is capable of describing a large variety of the open quantum dynamics, including quantum master equation, stochastic Liouville equation and hierarchical equation of motion. All the details about molecular structures and environments are contained in matrix $M$. 
$\rho_{\text{p}}$ and $\rho_{\text{c}}$ include the diagonal and off-diagonal elements of the density matrix in certain representation, respectively. By absorbing the coherence component, we can further obtain the reduced quantum dissipative equation of motion involving the populations only, through the Laplace transform
\begin{equation}
    \begin{split}
        |\tilde{\rho}_{\text{c}}(s)\rangle\rangle = (s-M_{\text{c}})^{-1}\left(M_{\text{cp}}|\tilde{\rho}_{\text{p}}(s)\rangle\rangle + |\tilde{\rho}_{\text{c}}(0)\rangle\rangle\right)
    \end{split}
\label{rhocs}
\end{equation}
where $|\tilde{\rho}_{\text{c}}(s)\rangle\rangle=\int_0^{\infty} e^{-st}|\rho_{\text{c}}(t)\rangle\rangle\text{d}t$. The inverse Laplace transform leads to
\begin{equation}
    \begin{split}
        |\rho_{\text{c}}(t)\rangle\rangle = e^{M_{\text{c}}t}|\rho_{\text{c}}(0)\rangle\rangle + \int_0^t e^{M_{\text{c}}(t-\tau)}M_{\text{cp}}|\rho_{\text{p}}(\tau)\rangle\rangle\text{d}\tau.
    \end{split}
\label{rhoc}
\end{equation}
Inserting Eq.(\ref{rhoc}) into Eq.(\ref{rhocs}) we find the reduced EOM for the population
\begin{equation}
    \begin{split}
        |\dot{\rho}_{\text{p}}(t)\rangle\rangle = M_{\text{p}}|\rho_{\text{p}}(t)\rangle\rangle + \int_0^t M_{\text{pc}}e^{M_{\text{c}}(t-\tau)}M_{\text{cp}}|\rho_{\text{p}}(\tau)\rangle\rangle\text{d}\tau.
    \end{split}
\end{equation}
where $M_{\text{p}}$ and $M_{\text{c}}$ are the population and coherence blocks of the full matrix $M$, whereas they couple with each other through the blocks $M_{\text{pc}}$ and $M_{\text{cp}}$. These two blocks characterize the quantum coherence effects, distinct from the classical dissipative equation of motion that will be clarified later on.

It is worth noting that the coherence decays faster than the population in many circumstances, for instance, the atomic coherence in lasing medium and the electronic coherence in conjugated molecules \cite{scully1967quantum,mukamel1997electronic,hestand2018expanded}. We can proceed via the approximation of the stationary coherence at the time of population:  $|\rho_{\text{c}}(t)\rangle\rangle=-M_{\text{c}}^{-1}M_{\text{cp}}|\rho_{\text{p}}(t)\rangle\rangle$. This results in the following form of the reduced EOM
\begin{equation}
    \begin{split}
        |\dot{\rho}_{\text{p}}\rangle\rangle = \left(M_{\text{p}} - M_{\text{pc}}M_{\text{c}}^{-1}M_{\text{cp}}\right)|\rho_{\text{p}}\rangle\rangle
    \end{split}
\label{rhopss}
\end{equation}
and for each element
\begin{equation}
    \begin{split}
        \dot{\rho}_{nn} = \sum_{m\neq n} \left(L_{nn,mm}\rho_{mm} - L_{mm,nn}\rho_{nn}\right)
    \end{split}
\label{rhonn}
\end{equation}
where $L\equiv M_{\text{p}} - M_{\text{pc}}M_{\text{c}}^{-1}M_{\text{cp}}$, and $L_{nn,nn}=-\sum_{m\neq n}L_{mm,nn}$ due to the probability conservation. Eq.(\ref{rhonn}) implies in general the non-vanishing net quantum flux from the state $m$ to the state $n$ with the quantum transition rate $L_{nn,mm}$, induces the curl quantum flux
\begin{equation}
    \begin{split}
        c_{mn} = L_{nn,mm}\rho_{mm} - \text{min}\left(L_{nn,mm}\rho_{mm},L_{mm,nn}\rho_{nn}\right)
    \end{split}
\label{curl}
\end{equation}
as illustrated in Fig.\ref{Schm}(a), and $c_{mn}\ge 0$. The rate matrix $t_{mn}=L_{nn,mm}\rho_{mm}$ collecting the transition rate between the states can be decomposed into the symmetric and asymmetric parts therein. The flux $c_{mn}$ renders the asymmetric part breaking the detailed balance while the symmetric part $\text{min}\left(L_{nn,mm}\rho_{mm},L_{mm,nn}\rho_{nn}\right)$ obeys the detailed balance. It has been proved in a math rigor that $c_{mn}$ is a superposition of loop fluxes being divergence free, if (1) $c_{mn}\ge 0$ for $m\neq n$ and $c_{nn}=0$, (2) $c_{mn}c_{nm}=0$ for $m\neq n$ and
(3) $\sum_m c_{mn} = \sum_n c_{mn}$ \cite{minping1982circulation,qian1979reversible}.

Obviously, the curl quantum flux in Eq.(\ref{curl}) satisfies the conditions listed above. Conditions (1) and (2) derive the unidirectional nature of the curl flux, resulting in the detailed-balance-violation. Condition (3) conserves the population. The quantum curl flux provides an intrinsic driving force to the irreversible dynamics of quantum particles, responsible for the out-of-equilibrium effects, such as the current and its fluctuation associated with the quantum transport.

Eq.(\ref{rhonn}) resembles the classical rate equation, but has completely different physics: it essentially involves the coherence while the classical description does not. Two prominent examples can be found in the Fr\"ohlich coherence of THz molecular vibrations and the exciton motion in photosynthesis \cite{zhang2019quantum,romero2014quantum}.

\section*{Relaxation and Fluctuation-Dissipation Relation}
A central problem in stochastic thermodynamics is the spontaneous fluctuations related to the relaxation dynamics governed by Eq.(\ref{curl}) which breaks the time reversibility. The latter is imprinted from the population transfer and decoherence that are important aspects in the spectroscopic study of the molecular relaxation and radiative processes. To elaborate this, we can send a weak probe field to drive the quantum system of interest away from the steady state. Let $V(t)$ denote the time-dependent interaction with the 

4 warnings
 probe field, the quantum dynamics reads
\begin{equation}
    \begin{split}
        |\dot{\rho}\rangle\rangle = \left[M - if(t)V_-(t)\right]|\rho\rangle\rangle
    \end{split}
\label{qmep}
\end{equation}
where $f(t)$ is classical external field. We denote the expectation of an observable in Liouville space as $\langle \Omega\rangle(t)=\langle\langle 1|\Omega_L|\rho(t)\rangle\rangle$. Solving the dynamics equation in the interaction picture and transforming back to the Schr\"odinger picture, we have $|\rho(t)\rangle\rangle = G(t){\cal \hat{T}}e^{-i\int_{-\infty}^t f(\tau) V_{\text{int},-} (\tau)\text{d}\tau} |\rho_{int,ss}\rangle\rangle$, where $\rho_{int,ss}$ is the steady state density matrix in the interaction picture. Up to the 1st-order expansion of the solution with respect to the coupling to the probe field, the response of the observable considered is given by
\begin{equation}
    \begin{split}
        \langle \delta \Omega\rangle (t) = -i\int_{-\infty}^t \text{d}t' \langle\langle 1|\Omega_L G(t-t')V_-(t')|\rho_{ss}\rangle\rangle f(t')
    \end{split}
\label{A1}
\end{equation}
where $\langle\delta\Omega\rangle(t)\equiv\langle \Omega\rangle(t)-\langle \Omega\rangle_{ss}=\langle\langle 1|\Omega_L|\rho(t)-\rho_{ss}\rangle\rangle$  is the expectation value of the response given the perturbation $f(t)V(t)$. Here, the operator $G(t)=e^{Mt}$ is the free propagator of the system in the absence of the probe field, $V_{\text{int},-}(t) = G^{-1}(t)V_-(t)G(t)$ and $|1\rangle\rangle=\sum_n |nn\rangle\rangle$ is the Liouville-space representation of the identity operator. The subscript $L$ ($R$) denotes the multiplication from the left (right), i.e., $\Omega_L|\rho\rangle\rangle\equiv |\Omega\rho\rangle\rangle$, ($\Omega_R|\rho\rangle\rangle\equiv |\rho \Omega\rangle\rangle$). The notation $V_-$ is defined by $V_-\equiv V_L-V_R$. The response function can be subsequently defined from Eq.(\ref{A1}). 


It follows straightforwardly that Eq.(\ref{A1}) reduces to the Kubo formula for thermal equilibrium $\rho_{ss}=Z^{-1}e^{-H/T}$. But we have to raise the question: {\it what will happen for the quantum systems driven far from equilibrium?} The thermodynamic reversibility is essentially broken, violating the detailed balance when driving the systems away from equilibrium. This must come into effect in the response of the system. To elaborate on this, we recast the density matrix into

\begin{equation}
    \begin{split}
        L_{nn,nn}\rho_{nn}^{ss} = -\sum_{m\neq n} \left[c_{mn} + \text{min}\left(L_{nn,mm}\rho_{mm}^{ss},L_{mm,nn}\rho_{nn}^{ss}\right)\right]
    \end{split}
\label{rhoss}
\end{equation}
via the curl flux for the steady state, and the coherence $\rho_{ml}^{ss}=\sum_n K_{ml,nn}\rho_{nn}^{ss},m\neq l$  
where $K=-M_{\text{c}}^{-1}M_{\text{cp}}$. From Eq.(\ref{rhoss}) and Eq.(\ref{A1}) and letting $W\equiv I+K$ it follows that
\begin{multline}
     R^{(1)}(t-t') = i \Big[\langle\langle 1|\Omega_L G(t-t')V_- W S_D|\rho_{ss}\rangle\rangle +\\ \langle\langle 1|\Omega_L G(t-t')V_ -W \text{V}_{ss}|\rho_{ss}\rangle\rangle\Big]
\label{R1tt}
\end{multline}
in a compact form for $t\ge t'$. $I$, $L_D$, $S_D$ and $\text{V}_{ss}$ are operators in the population subspace such that $I = \sum_n|nn\rangle\rangle\langle\langle nn|$ is the identity operator in the population subspace and $L_D^{-1} = \sum_n L_{nn,nn}^{-1}|nn\rangle\rangle\langle\langle nn|$. The operator $S_D = \sum_n S_{nn,nn}|nn\rangle\rangle\langle\langle nn|$, and $\text{V}_{ss} = \sum_n\sum_{k\neq n} \frac{c_{kn}}{L_{nn,nn}\rho_{nn}^{ss}}|nn\rangle\rangle\langle\langle nn|$,
where $S_{nn,nn}=L_{nn,nn}^{-1}\sum_{k\neq n}\text{min}\left(L_{nn,kk}\frac{\rho_{kk}^{ss}}{\rho_{nn}^{ss}},L_{kk,nn}\right)$. $\text{V}_{ss}$ consists of the reduced curl flux as a reminiscence of the classical flux \cite{fang2019nonequilibrium,wang2008potential}. Obviously, $S_D$ preserves the detailed balance because each term in the summation is symmetric under the exchange $k\leftrightarrow n$. As dictated by the curl flux, $\text{V}_{ss}$ measures the detailed-balance-violation. Taking Fourier transform of Eq.(\ref{R1tt}) we obtain the linear transmission of the system
\begin{multline}
    \mathrm{Im}R^{(1)}(\omega) = \mathrm{Re} \langle\langle 1|\Omega_L G(\omega)V_- W S_D|\rho_{ss}\rangle\rangle +\\  \mathrm{Re}\langle\langle 1|\Omega_L G(\omega)V_- W \mathrm{V}_{ss}|\rho_{ss}\rangle\rangle
\label{R1w}
\end{multline}
where $G(\omega)=\int_0^{\infty}G(t)e^{i\omega t}\text{d}t$ denotes the Green's function in the frequency domain. The {\it Nonequilibrium Fluctuation-Dissipation Relation} as given above conveys the information of how the system responses without the equilibrium statistics assumed. After choosing $\Omega=V$ and restraining to the thermal equilibrium, the coherence terms in the density matrix and the second term in the formula Eq.(\ref{R1w}) which is proportional to the nonequilibrium flux vanish. The formula Eq.(\ref{R1w}) written in terms of density matrix is
\begin{align}
    \mathrm{Im}R^{(1)}(\omega)&={\rm Tr} \{V_{int}(\omega)V_{int}(0)\rho_{int,ss}-V_{int}(\omega)\rho_{int,ss}V_{int}(0)\}\nonumber\\
    &=\int_0^\infty e^{i\omega t}\langle[V_{int}(t),V_{int}(0)]\rangle dt
\end{align}
 
This returns to the ordinary quantum FDR given by 
\begin{equation} 
\text{coth}(\hbar\omega/2k_B T)\text{Im}[R^{(1)}(\omega)]=S(\omega)+S(-\omega)\,,
\end{equation} 
where the fluctuation is given by $S(\omega)=\int_0^{\infty}e^{i\omega t}\langle V(t)V(0)\rangle\text{d}t$. 
The LHS of the equation represents the dissipation and the RHS encrypts the information of the fluctuation. In this case, the second term on the RHS of Eq.(\ref{R1w}) which is responsible for the quantum flux contribution for the out-of-equilibrium system is zero. The nonequilibrium FDR reveals that not only the steady-state fluctuation influences the relaxation process, the detailed-balance breaking sector, as quantified by the steady-state quantum flux between the states, also substantially contributes to the relaxation dynamics. This is a fundamental distinction from equilibrium systems where the relaxation is uniquely determined by the spontaneous fluctuations of the equilibrium quantum states. 

\begin{figure}[htb!]
\centering
{\includegraphics[width=0.4 \textwidth]{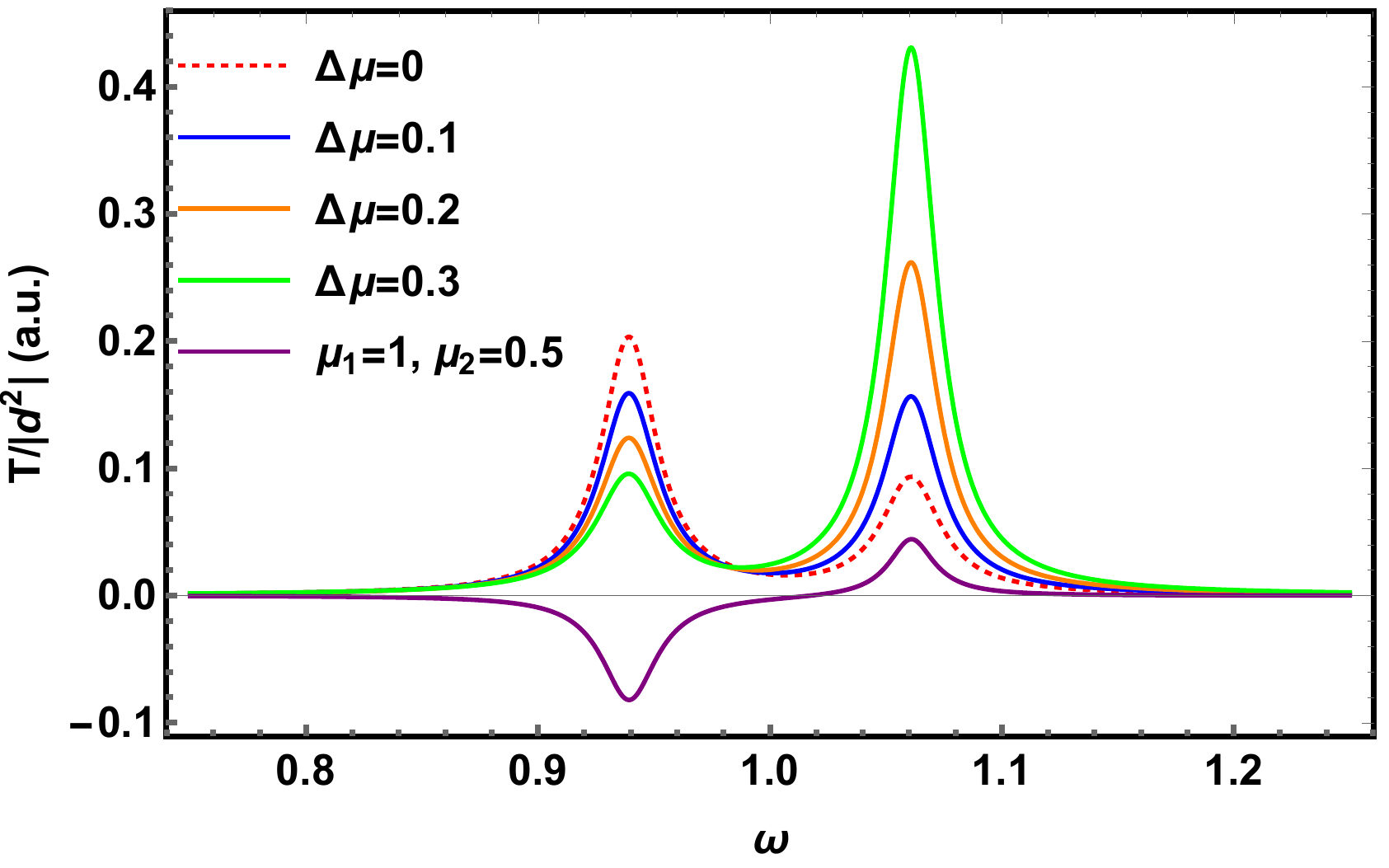}}
\quad
{\includegraphics[width=0.4 \textwidth]{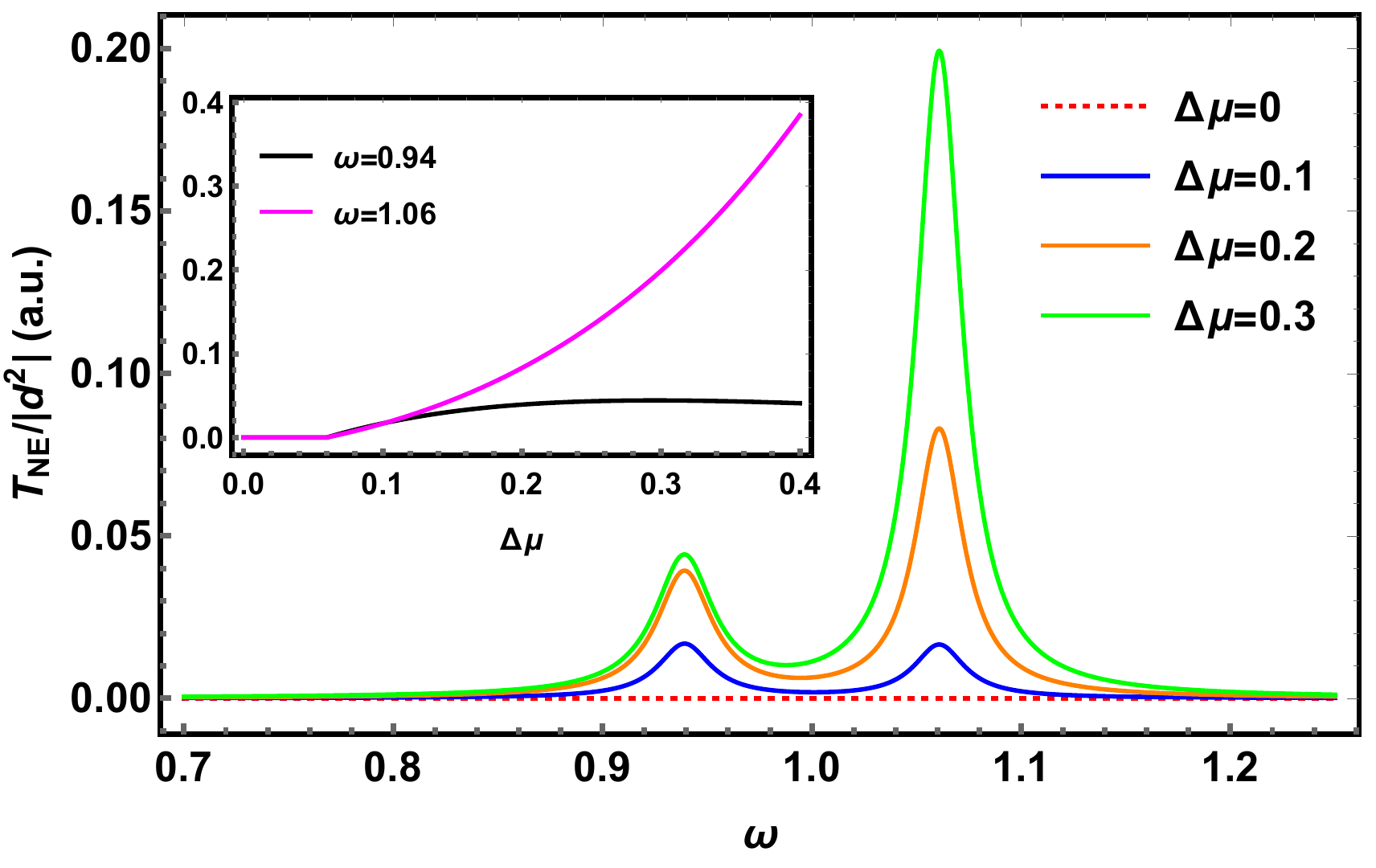}}
\quad
{\includegraphics[width=0.4 \textwidth]{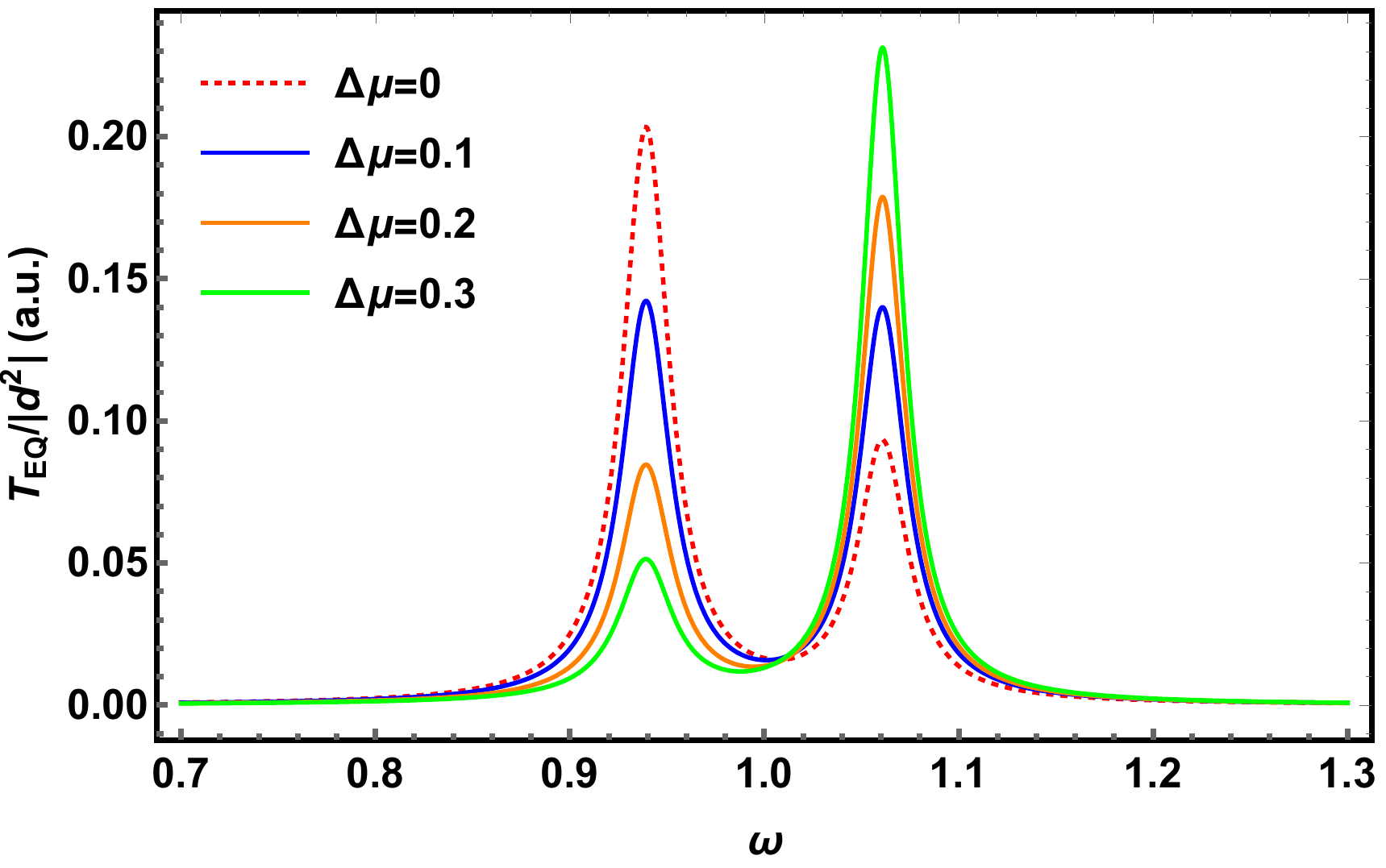}}
\caption{(a) Total transmission against the probe frequency at $\mu_{1(2)}=1\pm\Delta\mu$ and at $\mu_1=1$ and $\mu_2=0.5$. (b) Nonequilibrium contribution vs probe frequency. It shows the nonequilibrium component of the transmission $T_{\text{NE}}$ against the probe frequency at the chemical potential biases $\Delta \mu$. Various probe frequencies are shown in the small panel. (c) The equilibrium contribution vs probe frequency. The transmission due to the detailed-balance preserving contribution against the probe frequency under the same condition. Parameters are set to be $T_1=T_2=0.3$, $\Delta=0.01$, $\omega_{e_1g}=1.06,\ \omega_{e_2g}=0.94,\ \Gamma=0.02$.}  
\label{fig.1}
\end{figure}

It is worth noting the classical limit of the the nonequilibrium FDT given in Eq.(\ref{R1w}) as $K\rightarrow 0$. As such, the population equation in Eq.(\ref{rhonn}) reduces to the classical master equation in which $L_{nn,mm}$ measures the incoherent transition rates \cite{Esposito_PRL2016}. This classical version of FDT not only can describe the incoherent processes of quantum systems such as ballistic charge transfer, but may also be applied to a large variety of classical dissipative systems including protein dynamics and gene-expression control. This will be presented elsewhere.

While the present formalism of the FDT is for the finite-dimensional quantum systems, the nonequilibrium FDT for infinite-dimensional quantum systems may be an extensive issue in the forthcoming study. This is subject to an infinite-dimensional representation of the Lie group, yielding a classical limit dramatically different from the finite-dimensional systems \cite{agarwal1972fluctuation}.

\section*{Discussion}

Besides the 1st term on the right side that also appears in the FDT for equilibrium case, we observe an additional term breaking the thermodynamic reversibility as a result of the detailed-balance-violation. This detailed-balance-violation term does vanish whereas the 1st term still survives when the system returns to the thermal equilibrium, so that the conventional FDT is obtained. This is due to $\text{V}_{ss}=0$ given the curl flux $c_{mn}^{ss}=0$. Nevertheless, the FDR given in Eq.(\ref{R1w}) is general in the sense of involving the systems largely deviated from equilibrium, not restricted by the linear regime close to equilibrium. At the thermodynamic level the extra correlation governed by the curl flux in Eq.(\ref{R1w}) is therefore responsible for a deeper understanding of the quantum systems driven far from equilibrium. Moreover, the equilibrium component of the relaxation given by the 1st term is not completely exclusive to the nonequilibrium component. This can be seen from the fact that the curl flux depends on the population distribution. Notably, such correlation may be originated from the non-orthogonality between the driving forces from potential landscape and curl flux. This has been clarified for the classical stochastic systems involving Langevin noise \cite{fang2019nonequilibrium,wang2015landscape}.

In support of the universal nonequilibrium FDR in Eq.(\ref{R1w}), the charge transport in a molecular junction will be taken as an example to illustrate the above results.

\subsection*{Fluctuation-Dissipation Relation for Molecular Junctions}

To study the optical transmission of molecular junctions carrying electric current as a typical application of the universal FDT developed above, we adopt the simplest model for the open junctions where two coupled electronic states are subject to two electrodes having chemical potential bias. The Hamiltonian reads
\begin{equation}
    \begin{split}
        H_0 = \omega_{1} c_1^{\dagger}c_1 + \omega_{2} c_2^{\dagger}c_2 - \Delta(c_1^{\dagger}c_2+c_2^{\dagger}c_1) + U c_1^{\dagger}c_2^{\dagger}c_2 c_1
    \end{split}
\label{H0}
\end{equation}
where the $c$'s are the fermionic annihilation operators, i.e., $\{c_i,c_j^{\dagger}\}=\delta_{ij}$. $U$ quantifies the Coulomb interaction causing the blockade effect. Since we are interested in the strong blockade, single-electron transport dominates therein. We proceed via the polarization of the junction subject to a weak electric field $E(t)$, and the dipolar interaction reads $V(t)=-\hat \mu E(t)$ where $\hat \mu = \sum_{j=1}^2 \mu_{e_j g}|e_j\rangle\langle g| + \text{h.c.}$ and $\mu_{e_j g}$ is the matrix element of the transition dipole. The homodyne detection, as illustrated in Fig.\ref{Schm}(b), gives rise to the measurement of the transmission such that $T(\omega)=\text{Im}[E^*(\omega)P(\omega)]$ where $P(\omega)=\int P(t)e^{i\omega t}\text{d}t$ is the Fourier component of the polarization as a result of the grating and $P(t)$ gives the far-field dipolar radiation. Solving for the nonequilibrium density matrix, we find the linear response function for the junction
\begin{equation}
    R^{(1)}(\omega) = -i\langle\langle 1|V_LG(\omega)V_-|\rho_{ss}\rangle\rangle.
\label{R1MJ}
\end{equation}
This can be alternatively obtained by specifying the observable $\Omega$ in Eq.(\ref{A1}) to the dipole $V$. To evaluate the Green's function $G(\omega)$, we propagate the real-time dynamics of the electronic coherence $\rho_{g,e_j}$ following the equation
\begin{equation}
    \begin{split}
        \begin{pmatrix}
         \dot{\rho}_{g,e_1}\\[0.15cm]
         \dot{\rho}_{g,e_2}
        \end{pmatrix}
        =\begin{pmatrix}
          i\omega_{e_1,g}-\frac{\Gamma}{2}(\bar{f}_2+1) & -i\Delta\\[0.15cm]
          -i\Delta & i\omega_{e_2,g}-\frac{\Gamma}{2}(\bar{f}_1+1)
         \end{pmatrix}
         \begin{pmatrix}
          \rho_{g,e_1}\\[0.15cm]
          \rho_{g,e_2}
         \end{pmatrix}
    \end{split}
\label{rhoge}
\end{equation}
where $\bar{f}_j=[e^{(\omega_{e_jg}-\mu_j)/T}+1]^{-1}$ is the Fermi-Dirac distribution and $\mu_j$ denotes the chemical potential of the $j$-th electrode. $\Gamma$ is the rate of exchanging an electron between molecule and electrodes. Knowing the matrices $M_{\text{c}}$, $M_{\text{cp}}$, $K$ and $L$ for the junctions, we find the nonequilibrium contribution to the linear transmission spectrum
\begin{equation}
    \begin{split}
        T_{\text{NE}}(\omega) =\text{Im} R_{\text{NE}}(\omega)= |d|^2 J\text{Re}\left[{\cal G}(\omega)\right]
    \end{split}
\label{TNE}
\end{equation}
where $J$ denotes the magnitude of the curl flux in the junction and $\text{Im} R_{\text{NE}}(\omega)$ is the nonequilibrium contribution to the dissipation corresponding to the last term in Eq.\ref{R1w}. We have assumed $\mu_{e_j,g}\simeq d$ and ${\cal G}(\omega)$ is given in Supplemental Information. $T_{\text{NE}}(\omega)$ is closely related to quantum transport, evident by the flux involved in Eq.(\ref{TNE}). This will be elaborated later on. Because only a single loop exists in the three-level systems, so that
\begin{equation}
    \begin{split}
        J = L_{e_2e_2,e_1e_1}\rho_{e_1,e_1} - \text{min}\left(L_{e_2e_2,e_1e_1}\rho_{e_1,e_1},L_{e_1e_1,e_2e_2}\rho_{e_2,e_2}\right)
    \end{split}
\label{J}
\end{equation}
in the present model for molecular junctions. 

It turns out that the electric current inevitably results from the weighted superposition of the loop fluxes \cite{zhang2015landscape}. In the present model, the electric current is proportional to the flux itself, i.e., $i_e\propto \text{Im}\rho_{e_1,e_2}\propto J$, resulting from the Heisenberg equation of motion. 
This would make $T_{\text{NE}}(\omega)$ accessible in the experiments measuring the electron counting across the junctions.

We plot the transmission spectrum of the molecular junction in Fig.\ref{fig.1}, taking the parameters close to reality \cite{tao2010electron,nitzan2003electron}. Fig.\ref{fig.1}(a) shows that the junction absorbs the light differently for the excited modes when varying the voltage bias. When increasing the chemical potential bias, a prominent absorption of photons at higher excited state is observed, whereas the photon absorption by lower excited state is attenuated. This is attributed to the population imbalance coming from the detail-balance violation. To gain a deeper understanding of such nonequilibrium effect, we plot in Fig.\ref{fig.1}(b) the detail-balance-breaking part of the response function, and in Fig.\ref{fig.1}(c) the part of the response function keeping the detailed balance. Fig.\ref{fig.1}(b) illustrates the remarkable difference between the light absorption of the two electronic states, as increasing the voltage of the electrodes. Recalling the curl flux $J\propto L_{e_2e_2,e_1e_1} \rho_{e_1,e_1}-L_{e_1e_1,e_2e_2} \rho_{e_2,e_2}$ and the absorption intensity $\propto (\rho_{g,g}-\rho_{e_i,e_i})$, the detailed-balance violation essentially results in much larger population of higher excited state, giving rise to the intense light absorption by the higher excited state rather than the lower excited state. This can be further elaborated in the small panel of Fig.\ref{fig.1}(b), where the peak intensity against the bias is shown. Fig.\ref{fig.1}(c) shows that the equilibrium part of response function closely resembles the features of the full transmission spectrum as depicted in Fig.\ref{fig.1}(a), within the near-to-equilibrium regime ($\Delta\mu=0$ and $\Delta\mu=0.1$ for the dotted red and solid blue lines, respectively) where the reversibility is slightly broken. While in the far-from-equilibrium regime such that the curl flux is greatly enhanced ($\Delta\mu=0.2$ and $\Delta\mu=0.3$ for the orange and green lines, respectively), the nonequilibrium nature becomes significant in the linear transmission, by comparing Fig.\ref{fig.1}(b) with Fig.\ref{fig.1}(a).

Fig.\ref{fig.1}(b) and \ref{fig.1}(c) reveal very different physics: the nonequilibrium component of the relaxation is uniquely governed by the curl flux in Eq.(\ref{TNE}) while the equilibrium component mostly relies on the population distribution. Although the equilibrium component is not completely independent of the nonequilibrium one, the 2nd term in Eq.(\ref{R1w}) violating the reversibility characterizes the intrinsic nonequilibrium nature with the energy dissipation. This may lead to the experimental observations when measuring the electric current across the molecular junctions.

Lastly, one may notice in Fig.\ref{fig.1}(a) the dip in the transmission spectrum for a different set of parameters. An optical gain is thus indicated, resulting from the population inversion $ (\rho_{g,g}-\rho_{e_i,e_i})\propto$ absorption rate. This may reveal the lasing mechanism and cooperative radiation in molecular junctions, which will be presented elsewhere.

\section*{Conclusion}
In summary, we developed a universal relation between the fluctuations and the relaxation for nonequilibrium quantum systems. Using the curl flux theory, our work reveals the generic nature of the thermodynamic irreversibility such that the system relaxation consists of two distinct contributions: the spontaneous fluctuations showing the reversibility and the detailed balance breaking where the flux is predominately involved. This demonstrates completely different physics from the conventional FDT. The flux-induced contribution to the relaxation may lead to experimental observations when measuring the energy/charge transport via the electron or photon counting statistics that may bridge the gap between the quantum thermodynamics and the spectroscopy. As an example, we use the quantum master equation approach to derive the evolution of density matrix, obtaining the coherence dynamics and steady-state solution of molecular junctions. The molecule is subject to two electrodes that are modelled as fermionic reservoirs. The interaction with the reservoirs results in the exchange of electrons, responsible for the current across the junction. We derive the response function of nonequilibrium systems in Liouville space similar as the ultrafast spectroscopic work (Ref.\cite{mukamel1999principles,agarwalla2015coherent}), which is essential for incorporating the quantum curl flux to elaborate the roles of nonequilibriumness and transition pathways. For the detailed model calculations, we adopted the Born-Markov approximation to find the steady-state solution of the density matrix for molecular junction model and further propagate the electronic coherence. Our work offers new insights for understanding the quantum thermodynamics and the idea can be extended to nonlinear responses. Recent advance in multidimensional spectroscopy and thermodynamics provides a new platform to study the nonlinearity of the quantum systems driven far from equilibrium \cite{liphardt2002equilibrium,hoang2018experimental,zhang2018monitoring,dong2007quantum,galperin2017photonics,lomsadze2017frequency,agarwalla2015coherent,wang2021excitation}.

\section*{Acknowledgements}
Z.D.Z. gratefully acknowledges the support from grants ARPC-CityU new research initiative/infrastructure support from central (No. 9610505) and the Early Career Scheme from Hong Kong Research Grants Council (No. 21302721). We also thank Girish Agarwal and Wei Wu for the useful discussions.



\section*{Appendix}
In this supplementary material, we provide the details of obtaining the curl flux from the quantum master equation as well as the important steps in deriving the generalized fluctuation-dissipation relation in the model of molecular junctions. 

\subsection{Curl flux and open quantum dynamics}
The density matrix offers a powerful description for the dynamics of the quantum systems embedded in dense mediums. In a nonequilibrium process, the dynamics of the system can be decomposed into the detailed balance preserving component and the detailed balance breaking component. The detailed balance preserving component represents the equilibrium feature of the microscopic processes, while the detailed balance breaking is the intrinsic feature of the nonequilibriumness and its strength can be effectively represented by the curl flux decomposition \cite{fang2019nonequilibrium, zhang2014curl}. In this section, we will use the quantum master equation as a prominent example of dissipative quantum dynamics to elaborate the EOM in main text.

The environments consist of a group of harmonic oscillators having the Hamiltonian $H_B=\sum_m\sum_s v_s^{(m)}B_s^{(m),\dagger}B_s^{(m)}$ where $[B_s^{(m)},B_{s'}^{(n),\dagger}]=\delta_{mn}\delta_{ss'}$ and $m$ labels the individual environment. The system-environment interaction is assumed to be bilinear
\begin{equation}
    \begin{split}
        V(t) = \sum_{i,j}\sum_m\sum_s \lambda_{ij,s}^{(m)} A_{ij}(t)\left(B_s^{(m)}e^{-iv_s^{(m)}t} + B_s^{(m),\dagger}e^{iv_s^{(m)}t}\right)
    \end{split}
\label{V}
\end{equation}
in the interactive picture with $A_{ij}=|\psi_i\rangle\langle \psi_j|+|\psi_j\rangle\langle \psi_i|$ being the transition operator of the system. Suppose the environments are characterized by a smooth spectral density, the Born-Markoff approximation is applicable, allowing the 2nd-order truncation of the system-environment interaction
\begin{equation}
    \begin{split}
        \dot{\rho} = i[\rho,H] - e^{-iHt}\int_0^t dt' \text{Tr}_{\text{B}}[V(t),[V(t'),\rho\otimes\rho_{\text{B}}]] e^{iHt}
    \end{split}
\label{rho2}
\end{equation}
where the full density matrix of joint system-environment composition is  $\rho_T(t)=\rho(t)\otimes\rho_{\text{B}}(0)+\rho_c$ with the higher-order term $\rho_c$. $H$ is the Hamiltonian of the reduced quantum system of interest. Eq.(\ref{rho2}) neglects the higher-order term of the density matrix, assuming the back influence from the environments to the system has been neglected. This is because the environments contain a large number of particles so that their relaxation may be much faster than the system. Writing  $A_{ij}=A_{ij}^++A_{ij}^-$ where $A_{ij}^{+(-)}$ stands for the raising (lowering) operator between the $i$-th and $j$-th levels and adopting the rotating wave approximation (RWA), some algebra recasts Eq.(\ref{rho2}) into the QME
\begin{equation}
    \begin{split}
        \dot{\rho} = i[\rho,H] & + \sum_{i,j}\sum_{k,l}  \gamma_{ij}^{(+)}(\omega_{ij})\left(A_{ij}^+\rho A_{kl}^- - A_{kl}^- A_{ij}^+ \rho\right)\\[0.1cm]
        & \ \ + \gamma_{ij}^{(-)}(\omega_{ij})\left(A_{ij}^-\rho A_{kl}^+ - A_{kl}^+ A_{ij}^- \rho\right) + \text{h.c.}
    \end{split}
\label{qme}
\end{equation}
where $\gamma_{ij}^{(+)}(\omega_{ij})$ and $\gamma_{ij}^{(-)}(\omega_{ij})$ give the upward and downward transitions caused by environments, respectively. $\gamma_{ij}^{(+)}(\omega_{ij})/\gamma_{ij}^{(-)}(\omega_{ij})=e^{-\omega_{ij}/T}$ for thermal equilibrium in which $T$ is the bath temperature. In Liouville space, the QME in Eq.(\ref{qme}) falls into the matrix form of the EOM given in Eq.(\ref{qme1}). This will be beneficial for developing the real-time Green's function approach to study the quantum dynamics.

\subsection{Fluctuation-dissipation relation for molecular junctions}

To picturize the nonequilibrium FDT we have developed, elaborate efforts have to be further devoted to specific systems. We will hereafter apply the general formalism of FDT to the molecular junctions carrying the electric current, where the open junctions are modeled by two coupled electronic states subject to two electrodes having chemical potential bias \cite{thoss2018perspective}. In this section, we applies the QME approach to obtain the equations of motion for the reduced density matrix of the system and use the time-evolution operator derived from the QME to calculate the free propagator and the linear response function. The details of the  derivations are provided as follows.  

The molecular Hamiltonian takes the form of
\begin{equation}
    \begin{split}
        H_0 = \omega_g|0\rangle\langle 0| &+ \omega_1 c_1^{\dagger}c_1 + \omega_2 c_2^{\dagger}c_2 \\ &- \Delta(c_1^{\dagger}c_2 + c_2^{\dagger}c_1) + U c_1^{\dagger}c_2^{\dagger}c_2 c_1\,,
    \end{split}
\label{MH}
\end{equation}
where the $c$'s are the fermionic annihilation operators $\{c_n,c_m^{\dagger}\}=\delta_{nm}$. The fourth term in Eq.(\ref{MH}) describes the electron hopping between the two electronic states, and the last term comes from the Coulomb blockade. For the single electron transport, the molecular Hamiltonian may reduce to the effective Hamiltonian
\begin{equation}
    \begin{split}
        H_{\text{eff}} =\omega_g|g\rangle\langle g| + \omega_1 |e_1\rangle\langle e_1| + \omega_2 |e_2\rangle\langle e_2| - \Delta (|e_1\rangle\langle e_2|+|e_2\rangle\langle e_1|)
    \end{split}
\label{heff}
\end{equation}
by including the single-electron manifold only. Given the molecules are subject to two electrodes, the quantum master equation for the molecular junction is of the form
\begin{equation}
    \begin{split}
        \dot{\rho} = i[\rho,H_{\text{eff}}] + & \sum_{j=1}^2\frac{\Gamma_j}{2} \Big[\bar{f}_j\big(2|e_j\rangle\langle g|\rho|g\rangle\langle e_j| - \rho|g\rangle\langle g| - |g\rangle\langle g|\rho\big)\\[0.15cm]
        & + (1-\bar{f}_j)\big(2|g\rangle\langle e_j|\rho|e_j\rangle\langle g| - \rho|e_j\rangle\langle e_j| - |e_j\rangle\langle e_j|\rho\big)\Big]\,,
    \end{split}
\label{qmef}
\end{equation}
with the Fermi-Dirac distribution $\bar{f}_j=[e^{(\omega_{e_j g}-\mu_j)/T}+1]^{-1}$ where $\mu_j$ denotes the chemical potential of the $j$-th electrode. In what follows we assume $\omega_1>\omega_2$. We proceed via the polarization of the junction subject to a weak electric field $E(t)$, and the dipolar interaction reads $V(t)=-V E(t)$ where
\begin{equation}
    V = \sum_{j=1}^2 \mu_{e_j g}|e_j\rangle\langle g| + \text{h.c.}\,,
\label{dipole}
\end{equation}
and $\mu_{e_j g}$ is the matrix element of the electric dipole. The homodyne detection gives rise to the measurement of the transmission $\text{Im}[E^*(\omega)P(\omega)]$ where $P(\omega)=\int P(t)e^{i\omega t}\text{d}t$ is the Fourier component of the polarization as a result of the grating and $P(t)$ gives the far-field dipolar radiation. Quantum mechanically, the
far-field dipolar radiation is given by $\text{Tr}[V\rho(t)]$. Solving for the density matrix up to the 1st-order expansion with respect to molecule-field coupling, we find $P(\omega)=R^{(1)}(\omega)E(\omega)$ with the linear response function
\begin{equation}
    R^{(1)}(\omega) = -i\langle\langle 1|V_LG(\omega)V_-|\rho_{ss}\rangle\rangle
\label{R1MJ2}
\end{equation}
for the molecular junctions, by specifying the observable $\Omega$ to the dipole $V$ in Eq.(\ref{dipole}). To evaluate the real-time Green's function of the junction, we essentially propagate the coherence $\rho_{g e_j}$ which obeys the equation
\begin{equation}
    \begin{split}
        \begin{pmatrix}
         \dot{\rho}_{g,e_1}\\[0.15cm]
         \dot{\rho}_{g,e_2}
        \end{pmatrix}
        =\begin{pmatrix}
          i\omega_{e_1,g} - \frac{\Gamma}{2}(\bar{f}_2+1) & -i\Delta\\[0.15cm]
          -i\Delta & i\omega_{e_2,g} - \frac{\Gamma}{2}(\bar{f}_1+1)
         \end{pmatrix}
         \begin{pmatrix}
          \rho_{g,e_1}\\[0.15cm]
          \rho_{g,e_2}
         \end{pmatrix}
    \end{split}
\label{rhoge2}
\end{equation}
according to Eq.(\ref{qmef}) approximating $\Gamma_1\simeq \Gamma_2=\Gamma$. The propagator, which is essential in the calculation of the linear response function, is given by the exponentiation of the time-evolution matrix given in Eq.(\ref{rhoge2}). This gives rise to approximate propagator of the junction in the ``$ge$'' block given specifically as follows,
\begin{widetext}
\begin{equation}
    \begin{split}
       & G_{ge_1,ge_1}(t) = \frac{1}{2}\Bigg[\bigg(1 - \frac{\omega_{e_1,e_2}}{\sqrt{\omega_{e_1,e_2}^2+4\Delta^2}}\bigg)e^{(i\omega_--\gamma_-)t} + \bigg(1 + \frac{\omega_{e_1,e_2}}{\sqrt{\omega_{e_1,e_2}^2+4\Delta^2}}\bigg)e^{(i\omega_+-\gamma_+)t} \Bigg]\,,\\[0.3cm]
       & G_{ge_2,ge_2}(t) = \frac{1}{2}\Bigg[\bigg(1 + \frac{\omega_{e_1,e_2}}{\sqrt{\omega_{e_1,e_2}^2+4\Delta^2}}\bigg)e^{(i\omega_--\gamma_-)t} + \bigg(1 - \frac{\omega_{e_1,e_2}}{\sqrt{\omega_{e_1,e_2}^2+4\Delta^2}}\bigg)e^{(i\omega_+-\gamma_+)t} \Bigg]\,,\\[0.3cm]
       & G_{ge_1,ge_2}(t) = \frac{\Delta}{\sqrt{\omega_{e_1,e_2}^2+4\Delta^2}}\Big[e^{(i\omega_--\gamma_-)t} - e^{(i\omega_+-\gamma_+)t}\Big],\quad G_{ge_2,ge_1}(t) = G_{ge_1,ge_2}(t)\,,\\[0.3cm]
       & \gamma_{\pm} = \frac{\Gamma}{4}\bigg[2+\bar{f}_1+\bar{f}_2 \mp \frac{\omega_{e_1,e_2}(\bar{f}_1-\bar{f}_2)}{\sqrt{\omega_{e_1,e_2}^2+4\Delta^2}}\bigg]\ge 0\,, \quad G^{\{eg\}}(t)=G^{\{ge\},*}(t)\,,
    \end{split}
\label{Gge}
\end{equation}\end{widetext}
where $\omega_{\pm}=\frac{1}{2}\left(\omega_{e_1,g}+\omega_{e_2,g}\pm \sqrt{\omega_{e_1,e_2}^2+4\Delta^2}\right)$, $\omega_{e_i,g}=\omega_i-\omega_g$ is the excitation energy of the $i^{\mathrm{th}}$ molecule, and $\omega_{e_1,e_2}=\omega_1-\omega_2$ is the energy level bias between the two molecules. The crossing terms are identically zero, i.e. $G_{ge_i,e_jg}=G_{e_g,ge_j}=0$.

The linear response function is defined as
\begin{equation}
\begin{split}
            \text{Im}R^{(1)}(\omega)  = &\text{Re} \langle\langle 1|\Omega_L G(\omega)V_-(I+K)S_D|\rho_{ss}\rangle\rangle + \\ &\text{Re}\langle\langle 1|\Omega_L G(\omega)V_-(I+K)L_D^{-1}\text{V}_{ss}|\rho_{ss}\rangle\rangle\,,
\end{split}
\label{response}
\end{equation}
where $K=-M_{\text{c}}^{-1}M_{\text{cp}}$, $I = \sum_n|nn\rangle\rangle\langle\langle nn|$ is the identity matrix in the diagonal subspace, $L\equiv M_{\text{p}} - M_{\text{pc}}M_{\text{c}}^{-1}M_{\text{cp}}$ is the time evolution operator in the population subspace. The rest operators are given as follows, 
\begin{equation}
    \begin{split}
      & L_D^{-1} = \sum_n L_{nn,nn}^{-1}|nn\rangle\rangle\langle\langle nn|\,, \quad \text{V}_{ss} = \sum_n \sum_{k\neq n} \frac{c_{kn}}{\rho_{nn}^{ss}}|nn\rangle\rangle\langle\langle nn|\,,\\[0.15cm]
      & S_D = \sum_n \frac{1}{L_{nn,nn}}\sum_{k\neq n}\text{min}\left(L_{nn,kk}\frac{\rho_{kk}^{ss}}{\rho_{nn}^{ss}},L_{kk,nn}\right)|nn\rangle\rangle\langle\langle nn|\,,
    \end{split}
\end{equation}
where the definition of the curl flux $c_{nm}$ is given in Eq.~(\ref{curl}).

To calculate the response function, we need to find the expression of the operators $K,\ L,\ M_p,\ M_c$ and $M_{cp}$ from the QME. In the basis of $\{|gg\rrangle, |e_1e_1\rrangle, |e_2e_2\rrangle, |e_1e_2\rangle\rangle,\ |e_2e_1\rangle\rangle\}$, the matrices $M_{\text{c}}$ and $M_{\text{cp}}$ defined in Eq.~(\ref{qme1}) are given as follows,
\begin{equation}
    \begin{split}
       & M_{\text{c}} = \begin{pmatrix}
                         -i\omega_{e_1,e_2} - \frac{\Gamma}{2}(2-\bar{f}_1-\bar{f}_2) & 0\\[0.15cm]
                         0 & i\omega_{e_1,e_2} - \frac{\Gamma}{2}(2-\bar{f}_1-\bar{f}_2)
                        \end{pmatrix},\ \\[0.15cm]
        & M_{\text{cp}} = \begin{pmatrix}
                          0 & -i\Delta & i\Delta\\[0.15cm]
                          0 & i\Delta & -i\Delta
                         \end{pmatrix}    \,.
    \end{split}
\label{Mcoh}
\end{equation}
From the above information, we can find the representation of the operator $K\equiv-M_{\text{c}}^{-1}M_{\text{cp}}$ in the same subspace, 
\begin{equation}
    \begin{split}
        K = \begin{pmatrix}
             0 & -\frac{\Delta}{\omega_{e_1,e_2}-i\frac{\Gamma}{2}(2-\bar{f}_1-\bar{f}_2)} & \frac{\Delta}{\omega_{e_1,e_2}-i\frac{\Gamma}{2}(2-\bar{f}_1-\bar{f}_2)}\\[0.35cm]
             0 & -\frac{\Delta}{\omega_{e_1,e_2}+i\frac{\Gamma}{2}(2-\bar{f}_1-\bar{f}_2)} & \frac{\Delta}{\omega_{e_1,e_2}+i\frac{\Gamma}{2}(2-\bar{f}_1-\bar{f}_2)}
            \end{pmatrix}.
    \end{split}
\label{K}
\end{equation}
The QME in Liouville space can be written as $|\dot{\rho}_{\text{p}}\rangle\rangle = \left(M_{\text{p}} - M_{\text{pc}}M_{\text{c}}^{-1}M_{\text{cp}}\right)|\rho_{\text{p}}\rangle\rangle=L|\rho_{\text{p}}\rangle\rangle$, and the Liouville operator $L$ can be obtained from the Liouville operators $M_p$ and $M_{pc}$ given as follows, 
\begin{widetext}
\begin{equation}
    \begin{split}
         M_{\text{p}} = \begin{pmatrix}
                          -\Gamma(\bar{f}_1+\bar{f}_2) & \Gamma(1-\bar{f}_1) & \Gamma(1-\bar{f}_2)\\[0.15cm]
                           \Gamma\bar{f}_1 &  -\Gamma(1-\bar{f}_1) & 0\\
                           \Gamma\bar{f}_2 & 0 & -\Gamma(1-\bar{f}_2)
                         \end{pmatrix},   \text{ and }
        M_{\text{pc}}=\begin{pmatrix}
	     0 & 0\\
	     -i \Delta & i \Delta \\
	      i \Delta & -i \Delta 
	    \end{pmatrix}\,.
    \end{split}
\end{equation}
Therefore, the explicit form of the operator $L$ is given as follows,
\begin{equation}
    \begin{split}
         L = \begin{pmatrix}
                          -\Gamma(\bar{f}_1+\bar{f}_2) & \Gamma(1-\bar{f}_1) & \Gamma(1-\bar{f}_2)\\[0.15cm]
                           \Gamma \bar{f}_1 &  -\Gamma(1-\bar{f}_1)- \frac{\Delta^2 \Gamma (2-\bar{f}_1-\bar{f}_2)}{\omega_{e_1,e_2}^2+(\frac{\Gamma}{2}(2-\bar{f}_1-\bar{f}_2))^2}  & \frac{\Delta^2 \Gamma(2-\bar{f}_1-\bar{f}_2)}{\omega_{e_1,e_2}^2+(\frac{\Gamma}{2}(2-\bar{f}_1-\bar{f}_2))^2} \\
                           \Gamma\bar{f}_2 & \frac{\Delta^2 \Gamma (2-\bar{f}_1-\bar{f}_2)}{\omega_{e_1,e_2}^2+(\frac{\Gamma}{2}(2-\bar{f}_1-\bar{f}_2))^2}  & -\Gamma(1-\bar{f}_2)-\frac{\Delta^2 \Gamma (2-\bar{f}_1-\bar{f}_2)}{\omega_{e_1,e_2}^2+(\frac{\Gamma}{2}(2-\bar{f}_1-\bar{f}_2))^2} 
                \end{pmatrix}.
    \end{split}
    \label{L}
\end{equation}
\end{widetext}
From the explicit expression of $M_{\text{c}},\ M_{\text{cp}},\ K$ and $L$, we can obtain the expression for the nonequilibrium contribution to the response function defined as $R_{\text{NE}}''(\omega)=\text{Re}\langle\langle 1|V_L G(\omega)V_-(I+K)L_D^{-1}\text{V}_{ss}|\rho_{ss}\rangle\rangle =|\mu|^2J\mathrm{Re}[\mathcal{G}(\omega)]=|\mu|^2J\mathrm{Re}[\mathcal{G}^+(\omega)+\mathcal{G}^-(\omega)]$ in accordance to Eq.(\ref{R1MJ2}) by invoking the rotating wave approximation (RWA). The result is as follows,
\begin{equation}
    \begin{split}
        \mathcal{G}^{(+)}(\omega) = \bigg[ & \left(\frac{1}{L_{gg,gg}} - \frac{1+K_{e_1e_2,e_1e_1}}{L_{e_1e_1,e_1e_1}} - \frac{K_{e_1e_2,e_2e_2}}{L_{e_2e_2,e_2e_2}}\right)G_{e_1g,e_1g}(\omega)\\[0.25cm]
        & + \left(\frac{1}{L_{gg,gg}} - \frac{K_{e_2e_1,e_1e_1}}{L_{e_1e_1,e_1e_1}} - \frac{1+K_{e_2e_1,e_2e_2}}{L_{e_2e_2,e_2e_2}}\right)G_{e_2g,e_2g}(\omega)\\[0.25cm]
        & + \left(\frac{1}{L_{gg,gg}} - \frac{K_{e_2e_1,e_1e_1}}{L_{e_1e_1,e_1e_1}} - \frac{1+K_{e_2e_1,e_2e_2}}{L_{e_2e_2,e_2e_2}}\right)G_{e_1g,e_2g}(\omega)\\[0.25cm]
        & + \left(\frac{1}{L_{gg,gg}} - \frac{1+K_{e_1e_2,e_1e_1}}{L_{e_1e_1,e_1e_1}} - \frac{K_{e_1e_2,e_2e_2}}{L_{e_2e_2,e_2e_2}}\right)G_{e_2g,e_1g}(\omega)\bigg]\,,
    \end{split}
\label{RNE}
\end{equation}
where we have assumed $\mu_{e_1,g}\simeq\mu_{e_2,g}=\mu$ and that $J$ denotes the magnitude of the curl flux. Similarly, we can obtain $\mathcal{G}^-(\omega)(\omega)=\mathcal{G}(\omega)^{(+),*}(-\omega)$. Because only a single loop exits in three-level systems, $J=c_{g,e_1}=c_{e_1,e_2}=c_{e_2,g}$, we can write out the expression of the magnitude of the flux as follows,
\begin{equation}
    \begin{split}
        J = L_{e_2e_2,e_1e_1}\rho_{e_1,e_1}^{ss}-\text{min}(L_{e_2e_2,e_1e_1}\rho_{e_1,e_1}^{ss},L_{e_1e_1,e_2e_2}\rho_{e_2,e_2}^{ss})\,.
    \end{split}
\label{cf}
\end{equation}
With the above information, the propagators in frequency space can be obtained by taking Fourier transform to Eq.(\ref{Gge}),
\begin{equation}
    \begin{split}
        & G_{e_1g,e_1g}(\omega) = - \frac{\text{sin}^2\theta}{i(\omega-\omega_-)-\gamma_-} - \frac{\text{cos}^2\theta}{i(\omega-\omega_+)-\gamma_+},\\[0.25cm]
        & G_{e_2g,e_2g}(\omega) = - \frac{\text{cos}^2\theta}{i(\omega-\omega_-)-\gamma_-} - \frac{\text{sin}^2\theta}{i(\omega-\omega_+)-\gamma_+},\\[0.25cm]
        & G_{e_1g,e_2g}(\omega) = -\text{sin}\theta\text{cos}\theta\left[\frac{1}{i(\omega-\omega_-)-\gamma_-} - \frac{1}{i(\omega-\omega_+)-\gamma_+}\right],\\[0.25cm]
        & G_{e_1g,e_2g}(\omega) = -\text{sin}\theta\text{cos}\theta\left[\frac{1}{i(\omega-\omega_-)-\gamma_-} - \frac{1}{i(\omega-\omega_+)-\gamma_+}\right],\\[0.25cm]
        & \text{sin}2\theta = \frac{2\Delta}{\sqrt{\omega_{e_1,e_2}^2+4\Delta^2}},\ \text{cos}2\theta = \frac{\omega_{e_1,e_2}}{\sqrt{\omega_{e_1,e_2}^2+4\Delta^2}}\,.
    \end{split}
\end{equation}
From the homodyne detection $\text{Im}[E^*(\omega)P(\omega)]$ where $P(\omega)$ is the Fourier component of the far-field dipolar radiation and $P(\omega)=R^{(1)}(\omega)E(\omega)$, the positive peaks in the spectrum correspond to the enhancement of the transmission after the signal passing through the junction. Correspondingly, the negative peaks represent the absorption of the original signal.

%






\end{document}